# 3D STATISTICAL FACIAL RECONSTRUCTION

M. Berar
*LIS Laboratory*
*berar@lis.inpg.fr*

M. Desvignes
*LIS Laboratory*
*desvignes@inpg.fr*

G. Bailly
*ICP Laboratory*
*Bailly@icp.inpg.fr*

Y. Payan
*TIMC Laboratory*
*payan@imag.fr*

**Abstract**

*The aim of craniofacial reconstruction is to produce a likeness of a face from the skull. Few works in computerized assisted facial reconstruction have been done in the past, due to poor machine performances and data availability, and major works are manually reconstructions. In this paper, we present an approach to build 3D statistical models of the skull and the face with soft tissues from 3D CT scans. This statistical model is used by our reconstruction method to produce 3D soft tissues from the skull of one individual. Results on real data are presented and are promising.*

## 1. Introduction

Craniofacial reconstruction can be considered, when confronted with an unrecognizable corpse and where no other identification evidence is available. In such cases the skeletal remains are all that are available from which a picture of that person should be created. The aim of craniofacial reconstruction is then to produce a likeness of the face using the skeletalized remains. This reconstruction will hopefully provide a lead to enable a positive identification.

Several 3D manual methods for facial reconstruction have been developed and are currently used in practice. The reconstruction consists of modeling a face on a skull by use of clay and plasticine. However, manual reconstruction methods have several fundamental shortcomings, such as being highly subjective, time-consuming and requiring artistic talent. Computer-based methods were developed trying to provide an answer to the shortcomings of manual reconstruction.

Current computerized techniques either fit a template skin surface to a set of interactively placed virtual dowels on a 3D digitized model of the remaining skull [1] – [5] or deform a reference skull to the remaining skull based on crest lines (lines of maximal local curvature) [6], control data sets [7] or feature points [8] and apply an extrapolation of the calculated skull deformation to the skin surface associated to the reference skull. In both cases, the template skin or reference skull can either be a generic surface or a specific best look-alike according to the skull. However, the reconstruction is biased by the choice of reference skull or skin and the model of deformations between skull landmarks and between skull and skin is quite simple.

In this paper, a method to build a 3D statistical model of the skull and face is presented. With this model, 3D reconstruction of the face and soft tissues is available. The general idea is similar to [6-8] but uses a statistical shape model of the skull and the face for the reconstruction task, instead of an extrapolation of the deformation field. The deformation of the template skull is used to register the "remaining" skull in our model referential. A 3D-to-3D matching procedure delivers meshes of the skull and face with the same number of vertices. The matched vertices should refer to identical – in structural terms - facial and bony landmarks. Applied to several individuals, we are able to build a statistical model of the variability of the skull and the face. The reconstruction of the face is then resolved using the direct statistical relationship between skin and skull surface shapes given by the statistical model and can be seeing as a missing data problem.

This paper describes our approach for facial reconstruction. In section 2, we describe how normalized skull (and face) are obtained using a 3D-to-3D matching procedure. Section 3 presents the statistical model built upon the normalized faces and skulls and explains our facial reconstruction method. Finally, section 4 presents results and the guideline for furthers improvements.

## 2. Skull and Face Database

An entry (sample) in our database consists of a skull surface coupled with a skin surface. In case of facial reconstruction, only the skull surface is known. This surface is represented by a 3D mesh (vertices and triangles). In order to construct our statistical model, each skull or skin shape should share the same mesh structure with the same number of vertices. Then, all our meshes need to be registered in a subject-shared reference system. (figure 1, 3D deformed generic mesh i).

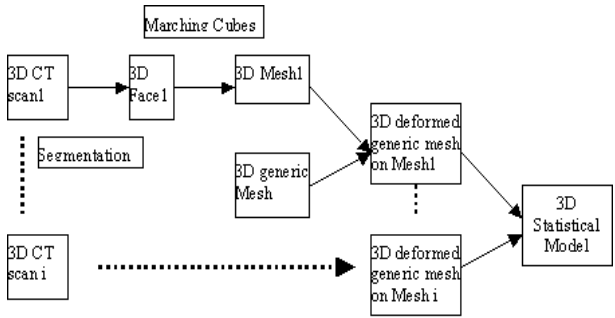

Figure 1 Building a 3D statistical model from 3D CT Scans. A 3D-to-3D matching procedure delivers meshes with the same number of vertices. The matched vertices should refer to identical – in structural terms - facial and bony landmarks.

In this system, the triangles for a region of the skull or the face are the same for all samples, while the variability of the position of the vertices will reflect the anatomical characteristics of each sample. The vertex of these shared meshes can be considered as semi-landmarks, i.e. as points that do not have names but that correspond across all the samples of a data set under a reasonable model of deformation from their common mean [9]. The shared meshes are obtained by matching generic meshes of the skull and the skin (see Figure 1 and 2) to several individual meshes using our 3D-to-3D matching algorithm.

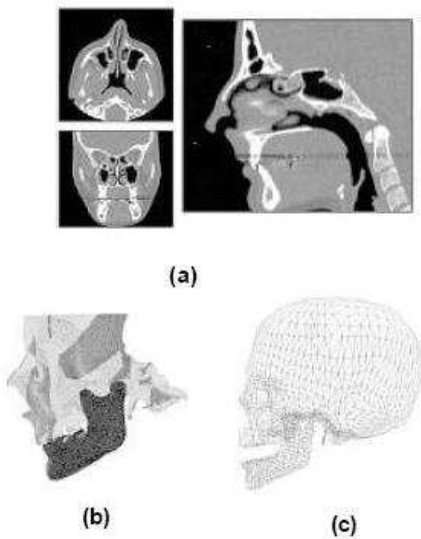

Figure 2. (a) 3D raw scan data (only coronal slices were collected; midsagital and axial have been reconstructed here by image processing), (b) shape reconstructed using the marching cube algorithm [10]; (c) generic mesh obtained from the Visible Woman Project®.

## 2.1. Acquisition and segmentation

Coronal CT slices (see Figure1.a) were collected for the partial skulls and faces of 9 subjects (helical scan with a 1-mm pitch and slices reconstructed every 0.31 mm or 0.48 mm). The Marching Cubes algorithm [10] has been implemented to reconstruct the skulls and the faces from CT slices on isosurfaces (see Figure1.b). The mandible and the skull are separated before the beginning of the matching process because our subjects have different mandible apertures. Patient-specific meshes for the skull, jaw and face have around 180000, 30000 and 22000 vertices. The respective generic meshes from the Visible Woman Project (for the skull and mandible) and from [11] (for the face) have 3473, 1100 (see Figure1.c) and 5828 vertices. Our 3D-to-3D matching algorithm described below is used to separate normalized meshes of these organs.

Table 1. Distance between transformed and target meshes.

| Distances (mm) | mean | Max |
|---|---|---|
| Mandible | 2 | 8 |
| Skull | 4 | 36 |
| Face | 1 | 5 |

A "symmetric matching" is used to obtain the mandible meshes (see section 2.2). Maximal distances are located on the teeth and on the coronoid process. The mean distances can be considered as the registration noise, due to the difference of density (see Table 1.). Teeth will not be part of our model, due to the frequent artefacts in CT scans.

Since these data were collected during regular medical exams and excitation of the brain volume is avoided if not necessary, nearly all the skull are partially scanned, and only 2 complete skull and face volume data were available. Therefore a partial mesh of each skull is first registered on the corresponding part of the 3D generic mesh. Symmetric matching (see Section 2.2) insures better registration, as the partial mesh and the original data have equivalent shapes. The rest of the whole patient mesh is transformed by this matching process with low noise as each vertex of the transformed partial mesh has an equivalent in the whole mesh. During this step, the cranial vault is (most of the time) inferred from the border of the skull, using the continuity of the transformation; it is not accurate and only the minimal fitting volumes will be used to build the statistical model. The maximal distances found in the resulting mesh are situated in the spikes beneath the skull, where the individual variability and the surface noise are large, due to segmentation errors.

Finally, face meshes are obtained using the same procedure. In this case, the maximal distances between generic and patient meshes are located around the eyes, that are part of the original data, but not part of the generic mesh.

## 2.2. 3D-to-3D matching

The basic principle of the 3D-to-3D matching procedure [12] consists of the deformation of the initial 3D space by a series of trilinear transformations $T_l$ applied to all vertices $q_i$ of elementary cubes (see Figure 3) of the generic mesh (source) towards the patient mesh (target):

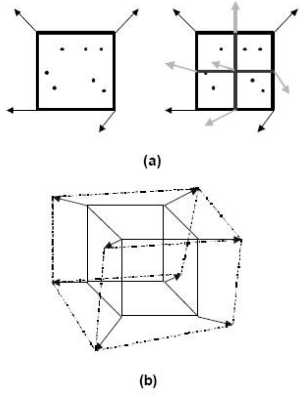

Figure 3. Applying a trilinear transformation to a cube. (a) 2D simplification of a subdivision into 4 elementary volumes of the original space and new transformations vectors; (b) elementary 3D transformation within a cube.

$$T_l(q_i, p) = \begin{bmatrix} p_{00} & p_{01} & & p_{07} \\ p_{10} & p_{11} & \cdots & p_{17} \\ p_{20} & p_{21} & & p_{27} \end{bmatrix} . [1 \ x_i \ y_i \ z_i \ x_iy_i \ y_iz_i \ z_ix_i \ x_iy_iz_i]^T$$

The parameters $p$ of each trilinear transformation $T_l$ are computed iteratively using the minimization of a cost function (see Eq.1 below). The elementary cubes are determined by iteratively subdividing the input space (see Figure 3) in order to minimize the distance between the 3D surfaces given by:

*(Eq. 1)*

$$\min_p \left[ \sum_{i=1; i \notin Paired(S_S, S_T)}^{card(S_S)} [d(T(t_i, p), S_S)]^2 + Rw. \sum_{k \in Paired(S_S, S_T)} [d(T(t_k, p), s_k)]^2 + P(p) \right]$$

where $S_S$ is the source surface to be adjusted to the set of points $\{t_i\}$ of the target surface $S_T$, $p$ are the parameters of the transformations $T$ (6 parameters of the initial rototranslation of the reference coordinate system and 3x8 parameters for each embedded trilinear transformation) applied to the set of points $\{s_i\}$ of $S_S$.

$P(p)$ is a regularization function that guaranties the continuity of the transformations at the limits of each subdivision of the 3D space. It allows larger deformations for smaller subdivisions.

In Eq.1, the first term deals with the distance between the points of the source mesh and the surface of the target mesh, considering the projection of each point onto the deformed surface.

The second term weighted by $R_w$ deals with point-to-point distance: a set of 3D feature points $\{t_k\}$ of the target surface $S_T$ are identified and paired with $\{s_k\}$ vertices of the source surface $S_S$. $R_w$ decreases with the number of iterations when the source and the target becomes close.

The minimization is performed using the Levenberg-Marquardt algorithm [13].

The problem of matching symmetry [14,15] is encountered, due to the difference of density between the source and target meshes (number of vertices respectively 30 and 70 times larger in the source meshes than in the target meshes). This problem can be observed using very different synthetic shapes. In Figure 4, our mismatched cone is well-matched considering the first distance but is flattened on one border of the sphere. Therefore, the minimization function is symmetrized by adding a term that computes also the distance of the target mesh to the transformed source mesh with the pseudo-inverse transform $T^{-1}$ in the following way:

$$\min_p \left[ \sum_{i=1; i \notin Paired(S_T)}^{card(S_T)} \left[d(T(t_i,p),S_S)\right]^2 + Rw. \sum_{k \in Paired(S_S, S_T)} \left[d(T(t_k,p),s_k)\right]^2 + P(p) \\ + \sum_{j=1; j \notin Paired(S_S)}^{card(S_S)} \left[d(T^{-1}(t_j,p),S_T)\right]^2 \right]$$

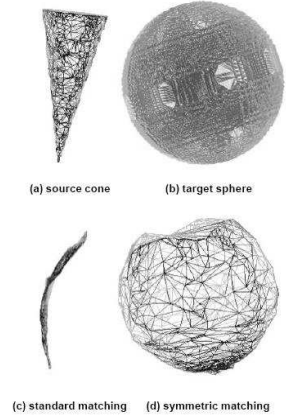

Figure 4. Matching a cone (a) to a sphere (b). (c) Mismatched cone using the single distance method. (d) Matched cone using the symmetric distance method

This new distance function solves this matching problem on synthetic cases (figure 4c) and on real skull and face meshes.

## 2.3. Statistical model

The statistical model is computed on the common part of the original data,. Minimum fitting volumes that depends on available CT scan data are extracted for the skull and the face (see Figure 5). Our nine matched skulls and faces are fitted on mean configuration of the skull using Procrustes normalization [16]. 7 degrees of freedom due to initial location and scale are retrieved by this fit (three due to translation along three axes, three due to rotations about three axes, one for scale adjustment). As the fitting is based on mean skull configuration of the skull, the relationships between each face and skull are conserved.

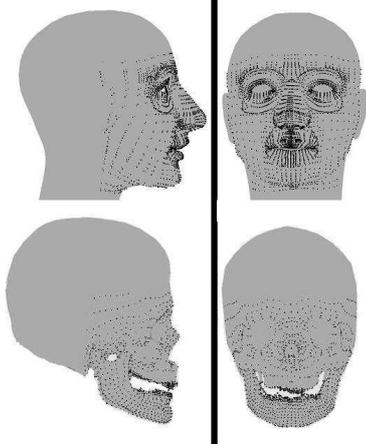

Figure 5. Minimum fitting volumes vertices.

A statistical model of the (partial) skull and the face is then built using Principal Component Analysis. The result of the PCA is a geometrically averaged facial template, which is computed together with a correlation-ranked set of modes of principal variations based on inter-subjects variations. Principal Component Analysis is an orthogonal basis transformation, where the new basis is found by diagonalizing the covariance matrix of a dataset.

Let $X_i = (x_{i1}, y_{i1}, ..., x_{in}, y_{in}) \in \mathbb{R}^{2n}$, be the locations of n vertives of the normalized meshes. Using PCA, we can write $T \approx \tilde{T} + \Phi b$, where $\tilde{T}$ is the mean mesh of the skull and face, $\Phi = (\phi_1 | ... | \phi_t)$ is a $(n+m) \times (n+m)$ matrix composed with the eigenvectors of the covariance matrix $S$ of the centered data and $b$ is a vector of $t$ dimension : $b = \Phi^t(T_i - \overline{T})$.

The dimension $t$ of the vector $b$ is the number of eigenvectors with the largest eigenvalues. In classical use of PCA, such as de-noising, $t$ is chosen by $\sum_{i=1}^{t} \lambda_i \geq 0.95 \sum_{i=1}^{m+n} \lambda_i$. The vector $b$ is then a good approximation for the original dataset and any of $n+m$ points can represented or retrieved with the $t_{t<n+m}$ values of the vector $b$ by $T \approx \tilde{T} + \Phi b$

## 3. Statistical Reconstruction

### 3.1. Reconstruction and Synthetic face

Every entry in the database is parameterized as a function of the statistical model. Instead of using a vector description of densely sampled points and skull landmarks, the entry is modelled as well as the sum of the geometrically averaged entry and a weighted linear combination of the modes of principal variation by $\tilde{T} + \Phi b$ with a fewer number of parameters. By altering the parameters $b$, new synthetic but valid skulls and faces, lying within the statistical boundaries of the are generated.

In our case, with 8 subjects, a total of 7 variations can be retrieved, but only the first three modes of variations are significant in term of represented variance.

The accuracy of this model is tested by reconstruction : for a given mesh, variation modes or parameter vector $b$ are computed by minimization of the distance between the true real mesh $T$ and reconstructed mesh from $b$ : $\tilde{T} + \Phi b$. The mean reconstruction errors (figure 6) for the last two modes are below the millimeter for samples of the learning database. So the reconstruction is quite accurate with sample in the learning database. Reconstruction error for a test sample i.e. a sample which is not in the learning database, is around 3 mm for the last four modes.

These two results demonstrate that this method is quite efficient but that the number of samples in the learning database is too small.

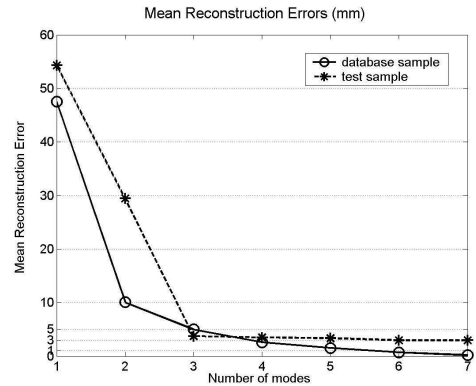

Figure 6. Mean reconstruction errors of the skull and face using an increasing number of mode

## 3.2. Missing Data Extension

The extension of the linear PCA model defined here is an elegant way to take into account spatial relations between landmarks and can also estimate the unknown part of the partially visible or occulted model [17].

Under this hypothesis, if some points (says $t=n$ points) are known, the remaining unknown points are determined using PCA. Without any approximations, we can write :

$$\begin{bmatrix} C_1 \\ \vdots \\ C_n \\ X_1 \\ \vdots \\ X_m \end{bmatrix} = \begin{bmatrix} \overline{C}_1 \\ \vdots \\ \overline{C}_n \\ \overline{X}_1 \\ \vdots \\ \overline{X}_m \end{bmatrix} + \begin{bmatrix} \Phi_{1,1} & \cdots & \Phi_{1,n+m} \\ \vdots & \ddots & \vdots \\ \Phi_{n+m,1} & \cdots & \Phi_{n+m,n+m} \end{bmatrix} \begin{bmatrix} b_1 \\ \vdots \\ b_n \\ b_{n+1} \\ \vdots \\ b_{n+m} \end{bmatrix}$$

This is a linear system with $n+m$ equations and unknowns that can not be resolved. Since PCA can represent the dataset with $t<n+m$ values, suppose $t=n$, the unknown vector $(b_1,\ldots,b_n,X_1,\ldots,X_m)$ in the following system. Notice, that if we choose $t<n$, the system become overdetermined and a least square method can be used to resolve the system :

$$\left\| \begin{bmatrix} C_1 - \overline{C}_1 \\ \vdots \\ C_n - \overline{C}_n \\ -\overline{C}_{n+1} \\ \vdots \\ -\overline{C}_{n+m} \end{bmatrix} - \begin{bmatrix} \Phi_{1,1} & \cdots & \Phi_{1,t} & 0 & \cdots & 0 \\ \vdots & & \vdots & \vdots & & \vdots \\ \Phi_{n,1} & \cdots & \Phi_{n,t} & 0 & \cdots & 0 \\ \Phi_{n+1,1} & \cdots & \Phi_{n+1,t} & -1 & 0 & \cdots & 0 \\ \vdots & & \vdots & 0 & \ddots & & \vdots \\ \vdots & & \vdots & \vdots & & \ddots & 0 \\ \Phi_{n+m,1} & \cdots & \Phi_{n+m,t} & 0 & \cdots & 0 & -1 \end{bmatrix} \begin{bmatrix} b_1 \\ \vdots \\ b_n \\ X_n \\ \vdots \\ X_{n+m} \end{bmatrix} \right\|^2$$

In this framework, a linear approximation of spatial relations between known and unknown points are explicitly determined from the eigenvectors of the covariance matrix.

## 3.3 Statistical Facial Reconstruction

Using the extension of the linear PCA defined above, the face of a subject can be reconstructed from his skull and from the statistical model defined previously. The known part (Ci) is the skull vertices, and the unknown part (Xi) is the face vertices.

A leave-one-out approach is used to test the accuracy of the facial reconstruction. The learning database is composed of all subjects minus one, which is the test sample. Every subject becomes the test sample in turn. Figure 7 gives the mean reconstruction error of the test sample which is not part of the learning base. It also gives the reconstruction error for the samples of the learning database.

In all cases, the global reconstruction is correct. The face is reconstructed with an accuracy of 0.5 mm for the samples in the learning database. Test sample is reconstructed with an mean accuracy of 7 mm. Clearly, these results show that the method is efficient but suffers from the size of our learning database.

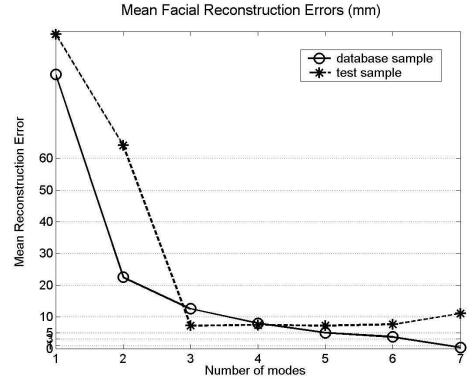

Figure 7. Mean Facial Reconstruction

Large errors are located on the nose and the boundary of the face (see figure 8). As the skull provides no information for the neck and very little information for the nose (as the vertices of the skull mesh are not sufficiently dense), the position of these points can not be accurately inferred. So the mean error is disturbed by these points. Another important factor is the number and the distribution of the vertices, which is not part of this study.

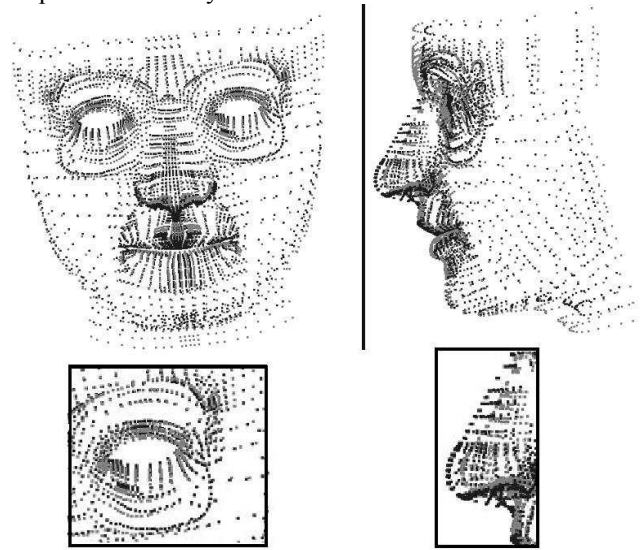

Figure 8. Reconstructed(black) and original (gray) face. The nose and the orbit are emphasized.

## 4. Conclusion

In this paper, a statistical model of the skull and face is proposed for 3D computerized facial reconstruction. The direct statistical relationship between the face and the skull included in the statistical model is used to reconstruct the missing data of the face when the skull is the only available information.

The main features of this approach are the idea of generic model to build a shared normalized space face and the statistical representation of spatial relationship between skull and face.

Results are visually correct and mean measured errors show that the method will be efficient for larger learning database, which is our future works.